\documentclass[twocolumn,aps,pra]{revtex4-1}
\usepackage{epsfig,amsmath,graphicx}
\usepackage{bbold}
\usepackage{colortbl}

\newcommand{\be}{\begin{equation}}
\newcommand{\ee}{\end{equation}}
\newcommand{\ba}{\begin{eqnarray}}
\newcommand{\ea}{\end{eqnarray}}

\begin{document}

\title{Quantum networks generation based on four-wave mixing}
\author{Yin Cai$^1$$^,$$^2$}

\author{Jingliang Feng$^2$}

\author{Hailong Wang$^2$}

\author{Giulia Ferrini$^1$}

\author{Xinye Xu$^2$}

\author{Jietai Jing$^2$}
\email{jtjing@phy.ecnu.edu.cn}

\author{Nicolas Treps$^1$}
\email{nicolas.treps@upmc.fr}

\affiliation{$^1$ Laboratoire Kastler Brossel, Sorbonne Universit\'e - UPMC, ENS, Coll\`ege de France, CNRS; 4 place Jussieu, 75252 Paris, France\\$^2$ State Key Laboratory of Precision Spectroscopy and Department of Physics, East China Normal University, Shanghai, 200062, China}

\date{\today}

\begin{abstract}

We present a scheme to realize versatile quantum networks by cascading several four-wave mixing (FWM) processes in warm rubidium vapors. FWM is an efficient $\chi^{(3)}$ nonlinear process, already used as a resource for multimode quantum state generation and which has been proved to be a promising candidate for applications to quantum information processing. We analyze theoretically the multimode output of cascaded FWM systems, derive its independent squeezed modes and show how, with phase controlled homodyne detection and digital post-processing, they can be turned into a versatile source of continuous variable cluster states.
\end{abstract}

\maketitle

\section{Introduction}
Generation of versatile quantum networks is one of the key features towards efficient and scalable quantum information processing. Recently their continuous variable implementation has raised a lot of interests \cite{Braunstein:2005wr}, in particular in optics where practical preparation and measurement protocols do exist, both at the theoretical and experimental level. The most promising achievements have been demonstrated using independent squeezed resources and a linear optical network \cite{Su:2007ts,Yukawa:2008iu}. More recently, proposals have emerged where different degrees of freedom of a single beam are used as the nodes of the network, such as spatial modes \cite{Armstrong:2012tt}, frequency modes \cite{Pysher:2011hn,Chen:2014jx}, or even temporal modes \cite{Yokoyama:2013jp}. In all these realizations, a given experimental setup corresponds to one quantum optical network. However, the specific structure of a quantum network depends on the mode basis on which it is interrogated, thus changing the detection system allows for on-demand network architecture. This has been applied in particular to ultra-fast optics \cite{Roslund:2013cb} where a pulse shaped homodyne detection is used to reveal any quantum network. In order to combine the flexibility of this mode dependent property with the simultaneous detection of all the modes, multi-pixel homodyne detection was introduced \cite{Armstrong:2012tt}, and it was shown that combined with phase control and signal post-processing it could be turned into a versatile source for quantum information processing\cite{Ferrini:2013cr}.

Here we propose a scheme based on four-wave mixing (FWM) in warm rubidium vapors to generate efficiently flexible quantum networks. A single FWM process can generate strong intensity-correlated twin beams \cite{McCormick:2007,Liu:2011,Qin:2012}, which has been proved to be a promising candidate in quantum information processing and has many applications such as quantum entangled imaging \cite{Boyer:2008}, realization of stopped light \cite{Camacho:2009} and high purity narrow-bandwidth single photons generation \cite{MacRae:2012}. Recently, it has been reported that by cascading two FWM processes, tunable delay of EPR entangled states \cite{Marino:2009}, low-noise amplification of an entangled state \cite{Pooser:2009}, realization of phase sensitive nonlinear interferometer \cite{Jing:2011,Kong:2013}, quantum mutual information \cite{Clark:2014} and three quantum correlated beams with stronger quantum correlations \cite{Qin:2014} can be realized experimentally. Inspired by these previous works we propose in the present work to cascade several FWM processes in which way we can turn this system into a controllable quantum network. We elaborate the theory of the optical quantum networks generated via cascading two and three FWM processes, calculating the covariance matrix and the eigenmodes of the processes from Bloch-Messiah decomposition \cite{Braunstein:2005fn}. We then study how cluster states can be measured using phase controlled homodyne detection and digital post-processing.

\section{Single FWM Process}

\begin{figure}[h]
%\centering
\includegraphics[width=8cm]{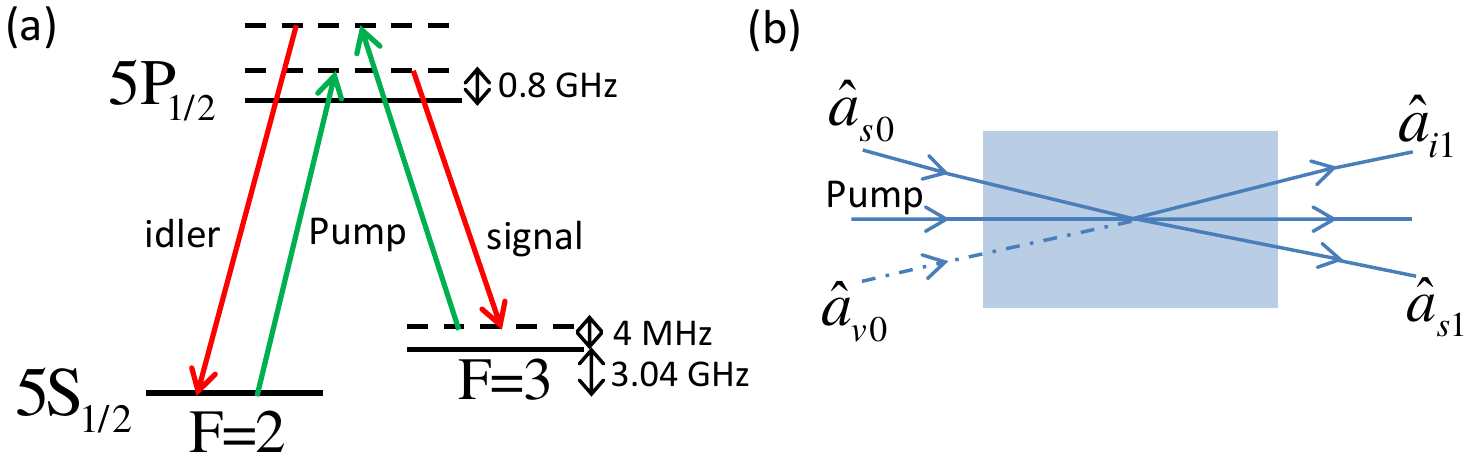}
\caption{(a) Energy level diagram for the FWM process. For experimental implementation the pump beam is tuned about 0.8 GHz to the blue of the D1 line of rubidium ($5S_{1/2}, F=2\rightarrow5P_{1/2}$, 795 nm) and the signal beam is red tuned about 3GHz to the pump beam. The two-photon detuning is about 4 MHz. (b) A single FWM process. $\hat{a}_{s0}$ is the coherent input and $\hat{a}_{v0}$ is the vacuum input. $\hat{a}_{s1}$ is the amplified signal beam and $\hat{a}_{i1}$ is the generated idler beam.} \label{fig principle}
\end{figure}

A single FWM process in Rb vapor is shown in Fig. \ref{fig principle}, where an intense pump beam and a much weaker signal beam are crossed in the center of the Rb vapor cell with a slight angle. During the process, the signal beam is amplified and a beam called idler beam is generated simultaneously. It propagates at the same pump-signal angle on the other side of the pump beam due to the phase-matching condition, having a frequency slightly shifted as compared to the signal beam. The input-output relation of the single FWM process is given by:

\begin{eqnarray}
\begin{aligned}
\hat{a}_{s1} & = & G\hat{a}_{s0}+g\hat{a}_{v0}^{\dagger} \\
\hat{a}_{i1} & = & g\hat{a}_{s0}^{\dagger}+G\hat{a}_{v0}
\end{aligned}
\end{eqnarray}
where G is the amplitude gain in the FWM process and $G^{2}-g^{2}=1$, $\hat{a}_{s0}$ is the coherent input and $\hat{a}_{v0}$ is the vacuum input. $\hat{a}_{s1}$ is the generated signal beam and $\hat{a}_{i1}$ is the generated idler beam, see \cite{Boyd:1992} for details.
Defining the amplitude and phase quadrature operators $\hat{X}=\hat{a}+\hat{a}^{\dagger}$ and $\hat{P}=i(\hat{a}^{\dagger}-\hat{a})$, the input-output relation can be re-written as:
\begin{equation}\label{eq:FWM1}
\left(
\begin{array}{c}
 \hat{X}_{\text{s1}} \\
 \hat{X}_{\text{i1}}
\end{array}
\right)=\left(
\begin{array}{cc}
 G & g \\
 g & G
\end{array}
\right)\left(
\begin{array}{c}
 \hat{X}_{\text{s0}} \\
 \hat{X}_{\text{v0}}
\end{array}
\right)\end{equation}

\begin{equation}\label{eq:FWM2}
\left(
\begin{array}{c}
 \hat{P}_{\text{s1}} \\
 \hat{P}_{\text{i1}}
\end{array}
\right)=\left(
\begin{array}{cc}
 G & -g \\
 -g & G
\end{array}
\right)\left(
\begin{array}{c}
 \hat{P}_{\text{s0}} \\
 \hat{P}_{\text{v0}}
\end{array}
\right)\end{equation}

We immediately see from this set of equations that the system does not couple $X$ and $P$ quadratures of the fields, which can thus be treated independently. Furthermore, input beams are vacuum or coherent states, and as the global transformation is symplectic the system retains gaussian statistic and can thus be fully characterized by its covariance matrix \cite{Braunstein:2005wr}. In our specific case, the covariance matrix is block diagonal:
\begin{equation}
C=\left(
\begin{array}{cc}
 C_{XX} & 0 \\
 0 & C_{PP}
\end{array}
\right)
\end{equation}
where, by definition, $C_{XX} =
\Big \langle \left(\begin{array}{c} \hat{X}_{\text{s1}} \\ \hat{X}_{\text{i1}} \end{array}\right)
\left(\begin{array}{c} \hat{X}_{\text{s1}} \\ \hat{X}_{\text{i1}} \end{array}\right)^T
\Big \rangle$, and the equivalent definition holds for $C_{PP}$. For coherent and vacuum input, the variances of input modes are normalized to one, and one obtains:
\begin{equation} \label{eq:CX}
{C_{XX}}=\left(\begin{array}{cc}
-1+2G^{2}& 2Gg\\
2Gg&-1+2G^{2} \end{array}
 \right)
\end{equation}
and
\begin{equation} \label{eq:CP}
{C_{PP}}=\left(\begin{array}{cc}
-1+2G^{2}& -2Gg\\
-2Gg&-1+2G^{2} \end{array}
 \right).
\end{equation}
${C}_{XX}$ and ${C}_{PP}$ are respectively the amplitude and phase quadrature parts of the covariance matrix of a single FWM process. The covariance matrix contains all the correlations between any two parties in the outputs. As the quantum state is pure, it is possible to diagonalize the covariance matrix to find the eigenmodes of the system, which are two uncorrelated squeezed modes, each one being a given linear combination of the output modes of the FWM process.  In this pure case $C_{PP}$ is simply the inverse of $C_{XX}$, so they share the same eigenmodes with inverse eigenvalues. We find that the eigenvalues of the $C_{XX}$ matrix are $\eta_{a1}=(G-g)^{2}$, $\eta_{b1}=(G+g)^{2}$ and the corresponding eigenmodes are $\hat X_{a1} = \frac{1}{\sqrt{2}}(\hat X_{s1} - \hat X_{i1})$ and $\hat X_{b1} = \frac{1}{\sqrt{2}}(\hat X_{s1} + \hat X_{i1})$. The first eigenmode is amplitude squeezed, while the second one is phase squeezed, which is the well known signature that, in a single stage FWM process, signal and idler beams are EPR correlated \cite{Marino:2009}.

It is important to stress here that each eigenmode of the covariance matrix is squeezed independently and diagonalization of the covariance matrix corresponds to a basis change from the output basis of FWM to squeezing basis. Even if this basis change can be difficult to be implemented experimentally, as output beams have different optical frequencies, it nevertheless remains a linear operation that reveals the underlying structure of the output state of the FWM process.

%One should also note that in this particular configuration the input signal beam (Stokes) and pump beam are coherent, while the input idler beam (anti-Stokes) is in vacuum state, and thus the system is phase insensitive\cite{Qin:2014}.

\section{Cascaded FWM Processes}
The above procedure can be readily applied to the more interesting multimode case, when one considers the multiple FWM processes, generating more than two output beams.  We study here three-mode asymmetrical and four-mode symmetrical structures, whose input-output relation is derived by successively applying the matrix corresponding to the single FWM process of Eq. (\ref{eq:FWM1}) and (\ref{eq:FWM2}).

\subsection{Asymmetrical structure: Double FWM Case}
 \begin{figure}[h]
%\centering
\includegraphics[width=8cm]{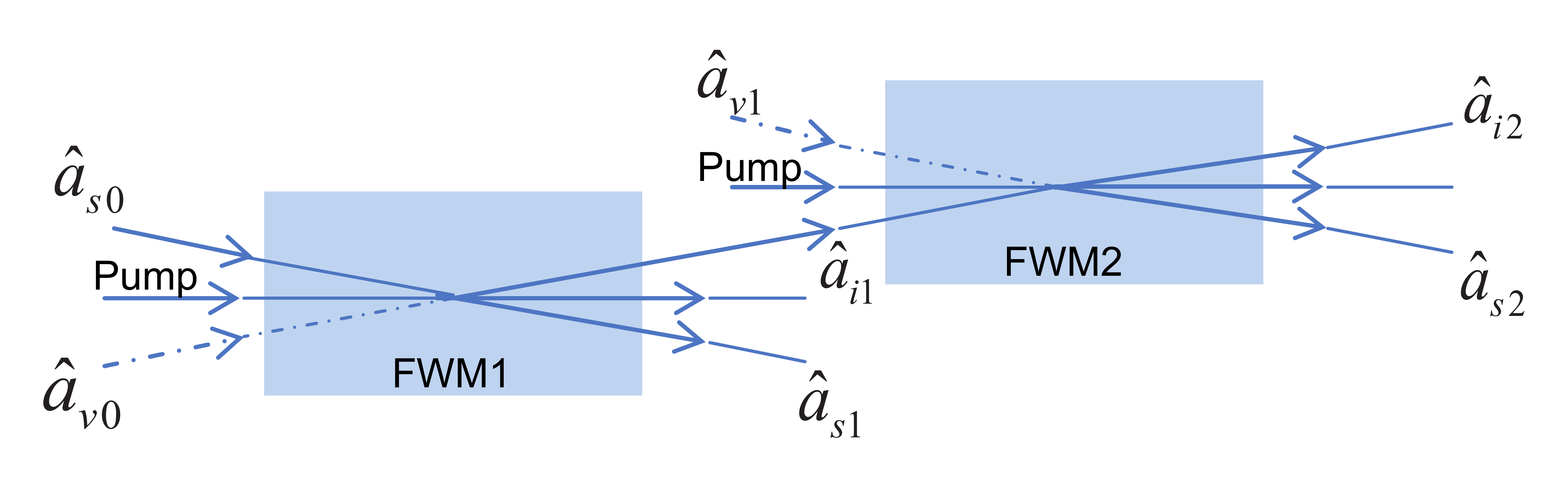}
\caption{Double stage structure of FWM Rb system. $\hat{a}_{s0}$ is the coherent input and $\hat{a}_{v0}$ is the vacuum input for the first FWM process. $\hat{a}_{s1}$ is the amplified signal beam and $\hat{a}_{i1}$ is the generated idler beam from the first FWM process. $\hat{a}_{v1}$ is the vacuum input for the second FWM process.  $\hat{a}_{s2}$ is the generated signal beam and $\hat{a}_{i2}$ is the amplified idler beam from the second FWM process.} \label{figdouble}
\end{figure}
We first consider the case where two FWM processes are cascaded. Without loss of generality, we take the idler beam from the first FWM process as the seed for the second FWM process, as described in Fig. \ref{figdouble}. The corresponding unitary transformation can be directly derived and writen:
\begin{eqnarray}
\begin{aligned}
\left(
\begin{array}{c}
 \hat{X}_{\text{s1}} \\
 \hat{X}_{\text{i2}} \\
 \hat{X}_{\text{s2}}
\end{array}
\right)& = U_{X_{3mode}}\left(
\begin{array}{c}
 \hat{X}_{\text{s0}} \\
 \hat{X}_{\text{v0}} \\
 \hat{X}_{\text{v1}}
\end{array}
\right)& \\
\left(
\begin{array}{c}
 \hat{P}_{\text{s1}} \\
 \hat{P}_{\text{i2}} \\
 \hat{P}_{\text{s2}}
\end{array}
\right)&=U_{P_{3mode}}\left(
\begin{array}{c}
 \hat{P}_{\text{s0}} \\
 \hat{P}_{\text{v0}} \\
 \hat{P}_{\text{v1}}
\end{array}
\right)&
\end{aligned}
\end{eqnarray}

where

\begin{eqnarray}
\begin{aligned}
U_{X_{3mode}}&=\left(
\begin{array}{ccc}
 G_1 & g_1 & 0 \\
 g_1 G_2 & G_1 G_2 & g_2 \\
 g_1 g_2 & g_2 G_1 & G_2
\end{array}
\right)&\\
U_{P_{3mode}}&=\left(
\begin{array}{ccc}
 G_1 & -g_1 & 0 \\
 -g_1 G_2 & G_1 G_2 & -g_2 \\
 g_1 g_2 & -g_2 G_1 & G_2
\end{array}
\right)&
\end{aligned}
\end{eqnarray}
Using the same procedure as for Eqs. (\ref{eq:CX}) and (\ref{eq:CP}) we can get the covariance matrix of the double stage FWM. It is still block diagonal, and for coherent or vacuum input states each block is given by:
\begin{equation}
C_{X_{3mode}}=U_{X_{3mode}}U_{X_{3mode}}^T
\end{equation}
\begin{equation}
C_{P_{3mode}}=U_{P_{3mode}}U_{P_{3mode}}^T\end{equation}

\begin{figure}[h]
%\centering
\includegraphics[width=8.5cm]{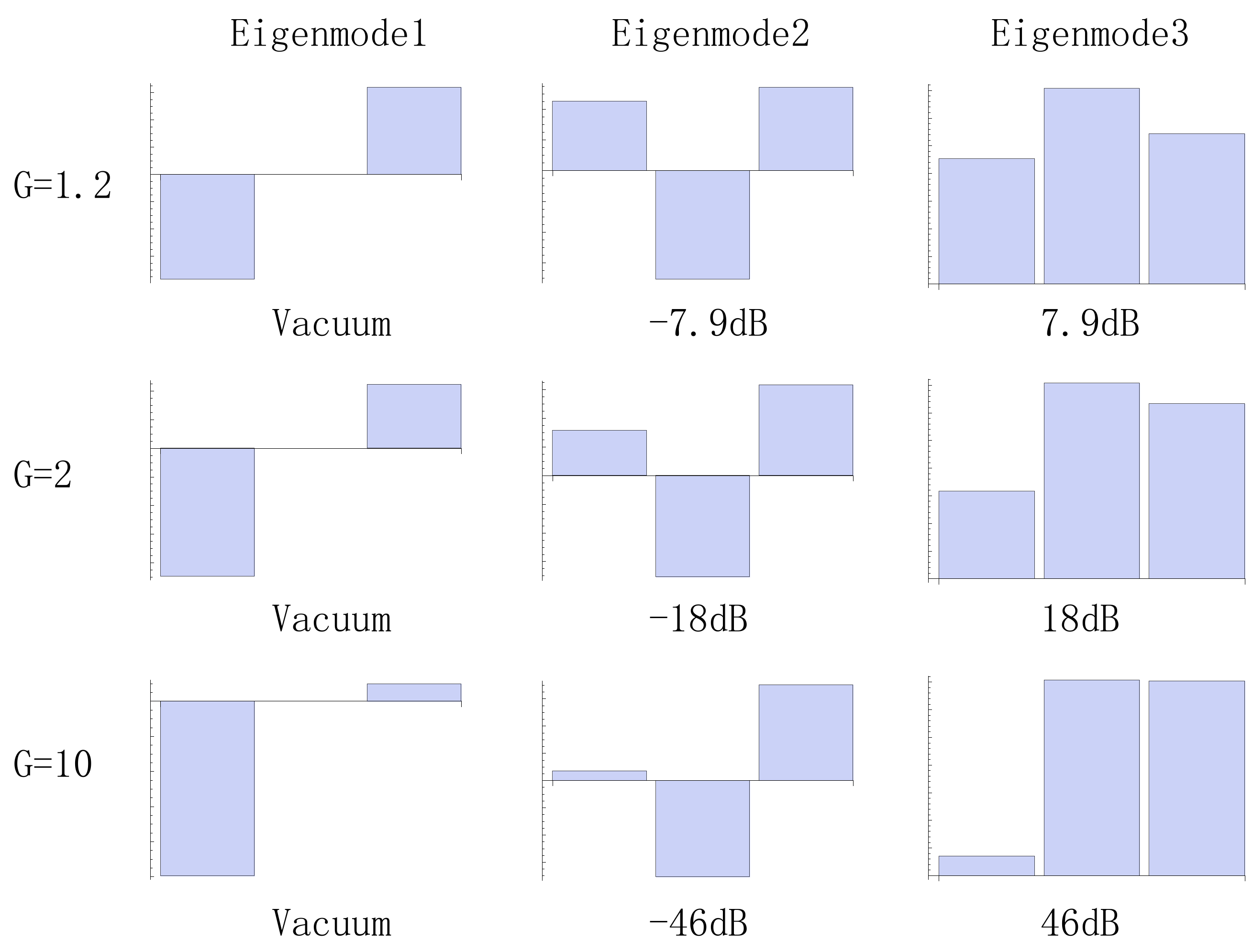}
\caption{Eigenmodes of the asymmetrical FWM cascade, decomposed in the FWM output mode basis, for three different gain values. For each graph, the bars represent the relative weight of modes $\hat a_{s1},\ \hat a_{i2},\ \hat a_{s2}$, respectively. Below are given the noise variances $\eta_{a3}$, $\eta_{b3}$ and $\eta_{c3}$ of the corresponding $\hat X$ quadrature. The state being pure, we see that eigenmode 3 shares the same squeezing as eigenmode 2 but on the phase quadrature.}
\label{threemodeFWM}
\end{figure}

We can now evaluate the eigenvalues and eigenmodes of these matrices. For the X quadrature, the eigenvalues of $U_{X_{3mode}}$ are:
\begin{equation}
\begin{aligned}
\eta_{a3}&=1&\\
\eta_{b3}&=-1+2\text{G}_{1}^2 \text{G}_{2}^2-2 \sqrt{\text{G}_{1}^2 \text{G}_{2}^2 \left(-1+\text{G}_{1}^2 \text{G}_{2}^2\right)}&\\
\eta_{c3}&=-1+2\text{G}_{1}^2 \text{G}_{2}^2+2\sqrt{\text{G}_{1}^2 \text{G}_{2}^2 \left(-1+\text{G}_{1}^2 \text{G}_{2}^2\right)}&
\end{aligned}
\end{equation}
Remarkably, one sees that one of the eigenvalues is equal to one, meaning that the system is composed of only two squeezed modes and one vacuum mode. This property can be extended if one generalizes this system to n-cell case in the similar asymmetrical way, there is always one vacuum mode. More expected, we also note that squeezing increases with gain, that eigenmode 2 and eigenmode 3 have the same squeezing but on different quadratures, and that both gains play an equivalent role and can be interchanged. The results for three different values of the gain, in the specific case where both processes share the same gain ($G_1 = G_2$) are shown in Fig. \ref{threemodeFWM}. We also show the shapes of the eigenmodes, i.e. their decomposition on the FWM output mode basis. The vacuum eigenmode appears to be composed only of modes 1 and 3 (i.e. $\hat a_{s1}$ and $\hat a_{s2}$), and tends to mode 1 when gain goes to infinity. This can be surprising, but it only reflects the fact that the noise of this mode becomes negligible compared to the two others when gain increases.

\subsection{Symmetrical structure: Triple FWM Case}
\begin{figure}[h]
%\centering
\includegraphics[width=8.5cm]{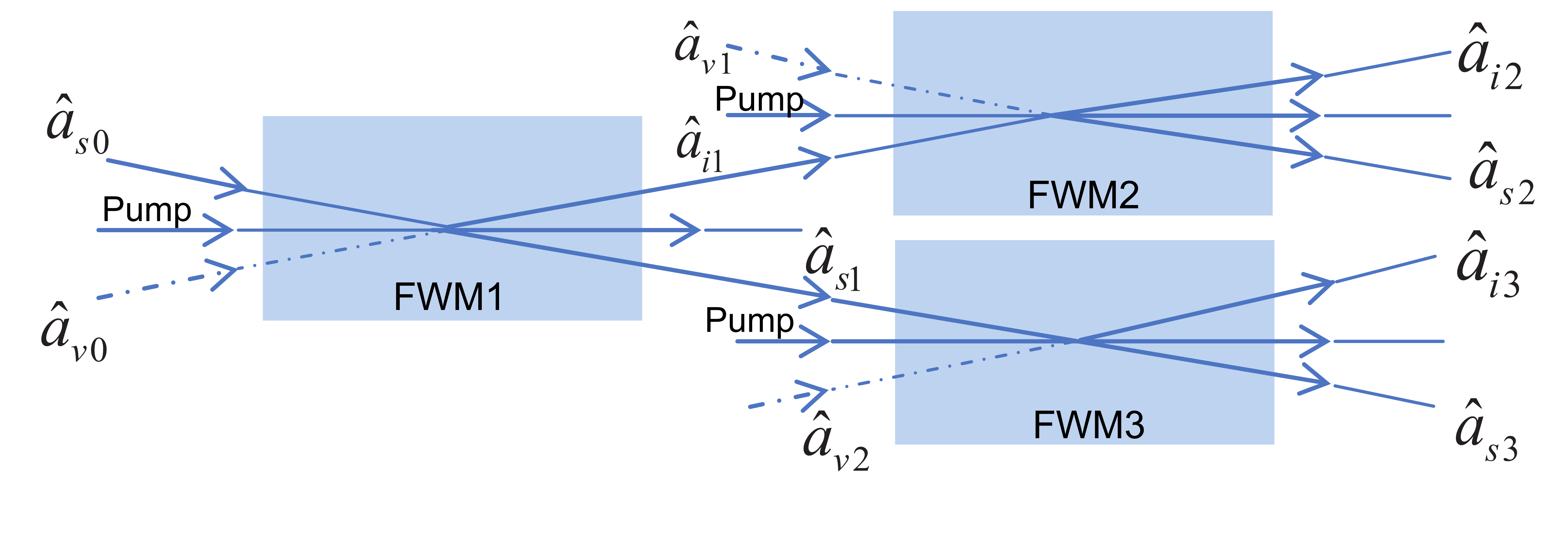}
\caption{Symmetrical structure of FWM Rb system. $\hat{a}_{s0}$ is the coherent input and $\hat{a}_{v0}$ is the vacuum input for the first FWM process. $\hat{a}_{s1}$ is the amplified signal beam and $\hat{a}_{i1}$ is the generated idler beam from the first FWM process. $\hat{a}_{v1}$ and $\hat{a}_{v2}$ are the vacuum inputs for the second and third FWM processes. $\hat{a}_{s2}$ is the generated signal beam and $\hat{a}_{i2}$ is the amplified idler beam from the second FWM process. $\hat{a}_{s3}$ is the amplified signal beam and $\hat{a}_{i3}$ is the generated idler beam from the third FWM process.}\label{figsymstage}
\end{figure}
We consider now the case of three cascaded FWM processes, where signal and idler of the first cell are used to seed each of the two other FWM processes, as shown in Fig. \ref{figsymstage}. For simplicity, we assume that all three FWM processes have the same gain value $G$. The evolution equations can be directly derived and lead to:
\begin{eqnarray}
\begin{aligned}
\left(
\begin{array}{c}
 \hat{X}_{\text{s3}} \\
 \hat{X}_{\text{i2}} \\
 \hat{X}_{\text{s2}} \\
 \hat{X}_{\text{i3}}
\end{array}
\right)&=U_{X_{4mode}}\left(
\begin{array}{c}
 \hat{X}_{\text{s0}} \\
 \hat{X}_{\text{v0}} \\
 \hat{X}_{\text{v1}} \\
 \hat{X}_{\text{v2}}
\end{array}
\right)&\\
\left(
\begin{array}{c}
 \hat{P}_{\text{s3}} \\
 \hat{P}_{\text{i2}} \\
 \hat{P}_{\text{s2}} \\
 \hat{P}_{\text{i3}}
\end{array}
\right)&=U_{P_{4mode}}\left(
\begin{array}{c}
 \hat{P}_{\text{s0}} \\
 \hat{P}_{\text{v0}} \\
 \hat{P}_{\text{v1}} \\
 \hat{P}_{\text{v2}}
\end{array}
\right)&
\end{aligned}
\end{eqnarray}

where,
\begin{eqnarray}
\begin{aligned}
U_{X_{4mode}}&=\left(
\begin{array}{cccc}
 G^2 & g G & 0 & g \\
 g G & G^2 & g & 0 \\
 g^2 & g G & G & 0 \\
 g G & g^2 & 0 & G
\end{array}
\right)&\\
U_{P_{4mode}}&=\left(
\begin{array}{cccc}
 G^2 & -g G & 0 & -g \\
 -g G & G^2 & -g & 0 \\
 g^2 & -g G & G & 0 \\
 -g G & g^2 & 0 & G
\end{array}
\right)&
\end{aligned}
\end{eqnarray}
%\begin{eqnarray}
%C_{X_3}=U_{X_3}U_{X_3}^T\nonumber\\
%C_{P_3}=U_{P_3}U_{P_3}^T\nonumber\\
%\end{eqnarray}

\begin{figure}[h]
%\centering
\includegraphics[width=9cm]{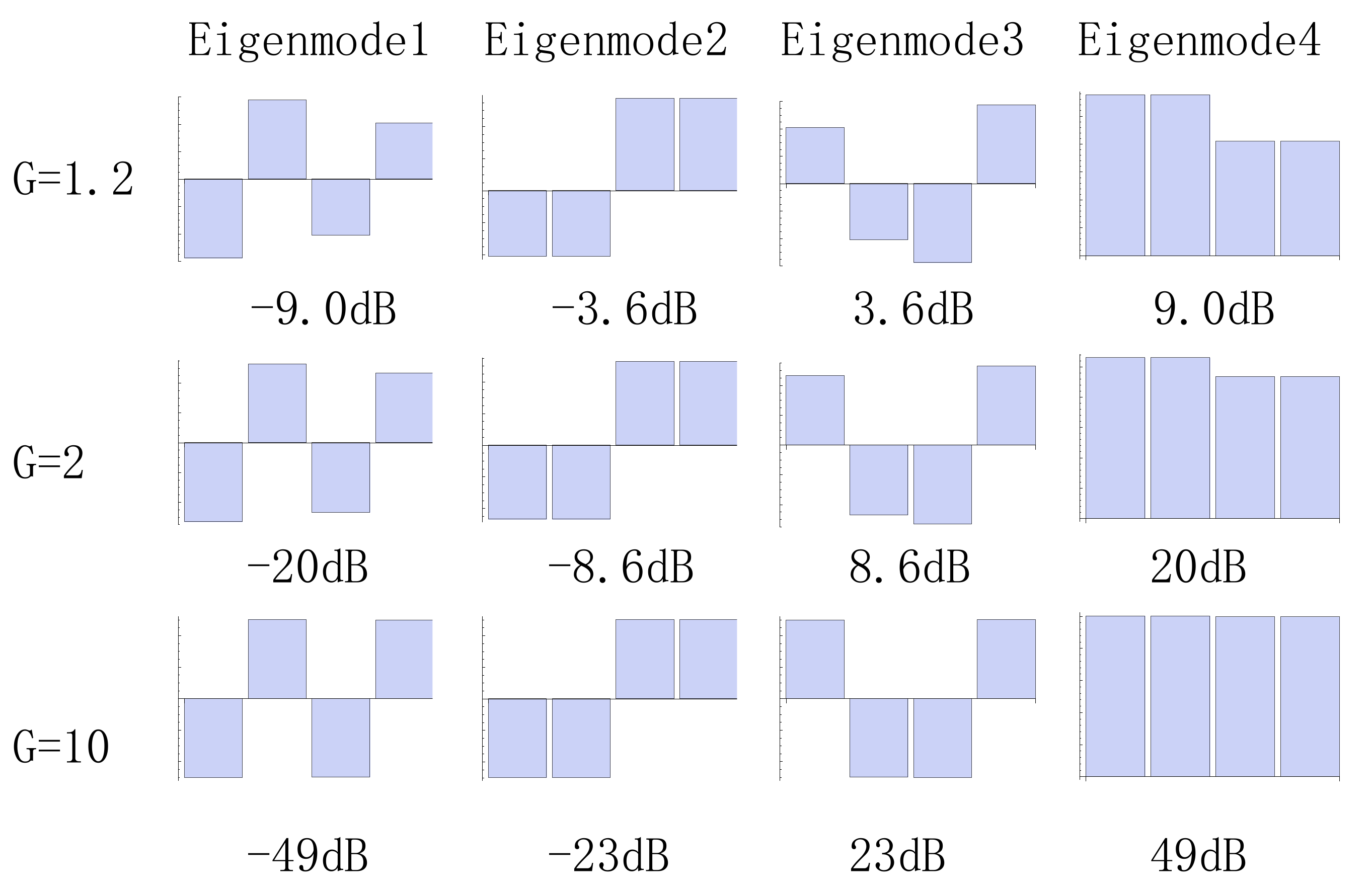}
\caption{Eigenmodes of the symmetrical 4-mode FWM cascade, decomposed in the FWM output modes basis, for three different gain values. For each graph, the bars represent the relative weight of modes $\hat a_{s3},\ \hat a_{i2},\ \hat a_{s2},\ \hat a_{i3}$, respectively. Below are given the noise variances of the corresponding $\hat X$ quadrature.}
\label{fourSqzFWM}
\end{figure}

No analytic expression of the eigenvalues can be simply given here, but for instance when G=1.2, we find for the X quadrature the following levels of squeezing $\{-9dB,-3.6dB,3.6dB,9dB\}$ (and opposite signs in the $P$ quadrature). This system is indeed composed of four independent squeezed modes, with two different squeezing values. Fig. \ref{fourSqzFWM} represents, similar as in the previous case, the mode shapes for three different values of the gain. As gain goes to infinity, we see that they tend to a perfectly symmetric decomposition, meaning that the output basis of FWM becomes mostly entangled then.

\section{cluster states}

We have shown in the previous section that the output states of different FWM processes were entangled states, whose underlying mode structure could be exactly calculated. We study here whether these outputs can be manipulated in order to generate cluster states, which are states of interest for quantum information processing. 

A cluster state is a specific multimode entangled state, defined through an adjacency matrix $V$ \cite{vanLoock:2010kt}. Let us call $\hat X_i^C$ and $\hat P_i^C$  the quadrature operators for the mode $\hat a_i^C $. The nullifier operators of the N-mode cluster states are defined by:
\begin{equation} \label{eq:nullifier}
\hat{\delta}_i =  \left( \hat{P}_{i}^{C} - \sum_{j} V_{ij} \cdot \hat{X}_{j}^{C} \right),
\end{equation}
Theoretically, a state is considered a cluster state of the adjacency matrix $V$ if and only if the variance of each nullifier approaches zero as the squeezing of the input modes approaches infinity, assuming that the cluster is built from a set of independently squeezed modes. Experimentally, one compares the variance of each nullifier to the corresponding standard quantum limit.

\begin{figure}[h]
%\centering
\includegraphics[width=8cm]{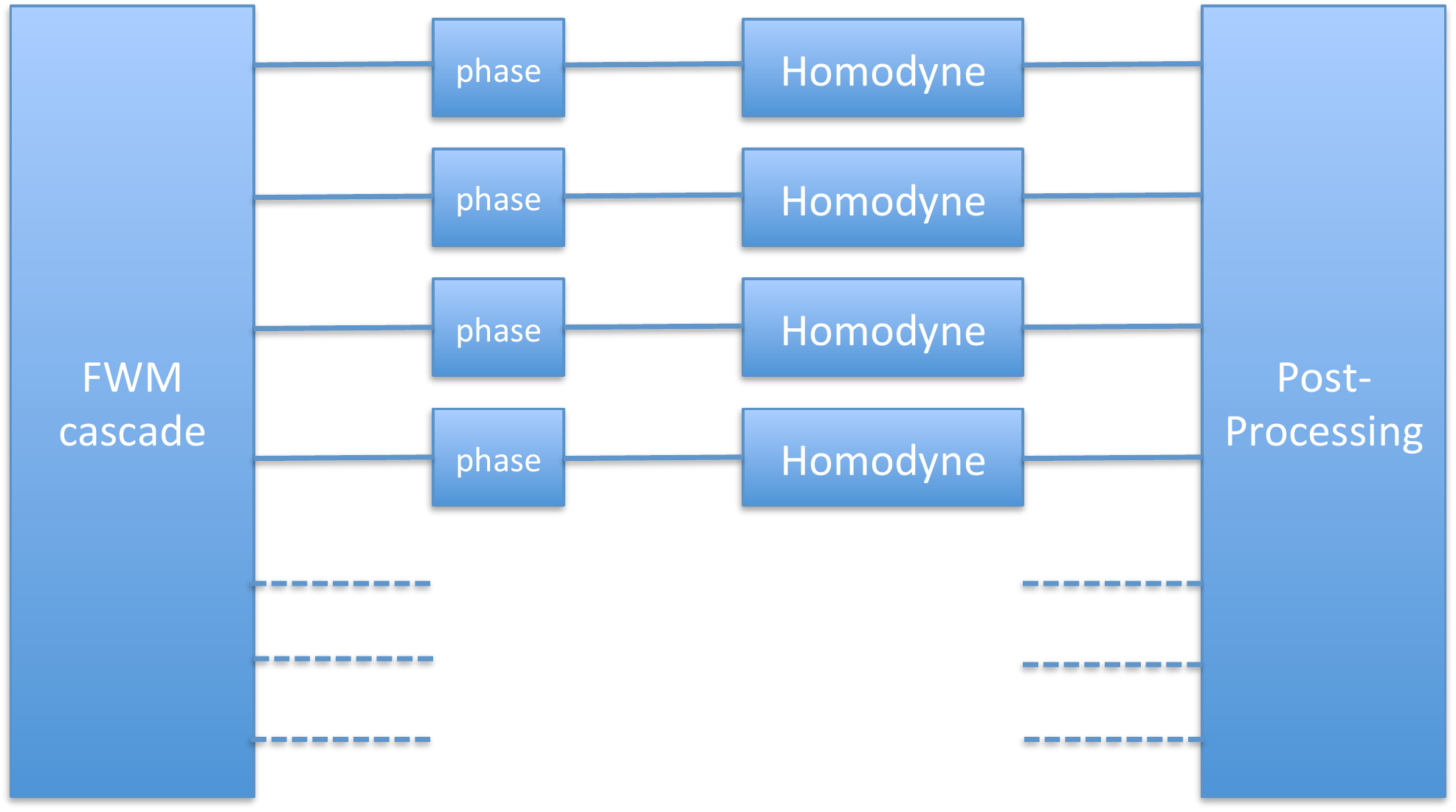}
\caption{Quantum networks can be constructed by applying phase controlled homodyne detections and post-processing the signals of the FWM outputs.}\label{FWMnetwork}
\end{figure}

It turns out that the output states of the FWM processes, as we have calculated in the previous sections, do not directly satisfy the cluster state criteria. However, it is still possible to derive cluster states when one can control the quadratures detected on each output mode (i.e. setting the phase of the homodyne detection local oscillator) and digitally post-process the data, as explained in \cite{Ferrini:2013cr}. To apply this theory to the present case, we first model the entangled state that one can produce with FWM, homodyne detection and post-processing, following the scheme of Fig. \ref{FWMnetwork}. First, to match the input of traditional cluster generation, we call $\hat a^\textrm{sqz}_i$ independent modes squeezed on the $P$ quadrature, with the squeezing values of the modeled FWM process (i.e. as displayed in Fig. \ref{threemodeFWM} and \ref{fourSqzFWM} for instance). Then we introduce the $U_{FWM}$ matrix so that $U_{FWM} \hat{\vec a}^\textrm{sqz}$, where $\hat{\vec a}^\textrm{sqz} = (\hat a^\textrm{sqz}_1, \hat a^\textrm{sqz}_2, \ldots)^T$, corresponds to the annihilation operators of the output modes of a given experimental setup. One can write:

\begin{equation}
U_{FWM}=U_{0}P_{sqz}
\end{equation}
where $P_{sqz}$ is a diagonal matrix which rotates the squeezing quadrature so that they match the results of previous sections and $U_0$ is a basis change from the squeezing basis to the output basis of the FWM setup, where homodyne detection is performed. With this convention, $U_0$ can be directly linked to the basis change matrices calculated in previous sections. Indeed, if for a given FWM process we call $D = diag(\eta_1, \eta_2, \ldots)$ the diagonal matrix composed of the eigenvalues of the process, then by definition the covariance matrix can be decomposed as $C_{Xnmodes} = U_0 D U_0^T$.
Then, the total transformation can be written as:
\begin{equation}
U_{total}=O_{post}P_{homo}U_{FWM}
\end{equation}
where $P_{homo}$ is a diagonal matrix that sets the quadrature measured by each homodyne detection, and $O_{post}$ is an orthogonal matrix describing post-processing by computer on the photocurrents measured by the homodyne detections.

We now compare this transformation to a given cluster state matrix $U_V$. Traditionally, $U_V$ is a matrix that moves from $p$ squeezed modes to cluster state modes, with $V$ the cluster adjacency matrix\cite{vanLoock2007}. Thus, the system is equivalent to a cluster state if one can find experimental parameters such that:
\begin{equation}
U_{V}=O_{post}P_{homo}U_{0}P_{sqz}
\end{equation}
In practice, it is possible to act on the gains of the different FWM processes, the local oscillators phases $P_{homo}$, and the post-processing operations $O_{post}$ to make the system achieve the transformation $U_V$ of the clusters state. According to \cite{Ferrini:2013cr}, defining $U'_V=U_{V}R^\dagger$ with $R=U_{0}P_{sqz}$, this problem has a solution if and only if $U_{V}^{'T}U_{V}^{'}$ is a diagonal matrix. Equivalently, if and only if one can write:
\begin{equation}\label{eq:criteria}
P_{homo}^2=U_{V}^{'T}U_{V}^{'}.\end{equation}
 In that case, one finds that $O_{post}$ is given by:
\begin{equation}
O_{post}=U'_{V}P^{-1}_{homo}.
\end{equation}
Using this formalism, it is thus possible to exploit the entanglement naturally generated by the cascaded FWM processes in order to generate cluster states. We will see in the following how it is possible to optimize the different experimental parameters to achieve some specific clusters.

\section{optimizations and solutions}

For a given cluster state specified by its adjacency matrix $V$, one can directly check whether using proper phases for homodyne detection ($P_{homo}$) and post-processing with a computer ($O_{post}$) it is possible to realize the cluster state $U_V$. Furthermore, one can demonstrate that if $U_V$ is a unitary matrix that leads to a cluster defined by $V$, then for any arbitrary orthogonal matrix  $O$, $U_V O$ leads to the same cluster state \cite{NewGiulia}. Thus, it is possible to run a searching algorithm to find an $O$ matrix that allows to satisfy our criteria of cluster generation. In practice, and as this is numerical calculation, we never find the exact equality in equation (\ref{eq:criteria}), thus we run an evolutionary algorithm \cite{Jonathan} leading to the matrix which is the closest to a diagonal one, then keep only the diagonal terms (re-normalized to one) to define the $P_{homo}$ matrix, and finally calculate the values of the nullifiers. This is the optimization procedure which is applied to find the results below.

\subsection{three-mode cascaded FWM}
We first start with the three-mode cascaded FWM process, which we have demonstrated is composed of only two squeezed modes and one vacuum mode. There are only two possible cluster graphs in that case, and as an exemple we study here only the possibility to generate a linear cluster state. The corresponding $U_V$ matrix can be found in \cite{Yukawa:2008iu}. We choose gains values $G_1=G_2=1.2$ as they give realistic experimental squeezing values. Performing the optimization with an evolutionary algorithm, we find solutions for the three-mode linear cluster state (matrix values given in the appendix). The normalized nullifiers are $\{0.22,0.16,0.94\}$, all below the shot noise limit, meaning that the 3-mode linear cluster state can be generated by the structure of the FWM. But there is no feasible solution when $G_1=G_2=2$, or for higher values of the gain. This can be surprising, but is directly linked to the mode structure at the output of the asymmetrical FWM, where one eigenmode is vacuum, and is getting closer to the first mode while gain increases, making it impossible to be transferred into a cluster state by post-processing. The nullifiers values are summarized in Table \ref{threemodenullifier}.

\begin{table}[h]
\centering
\begin{tabular}{|c|c|c|c|} \hline
FWM gain & nullifier 1 &nullifier 2 & nullifier 3\\ \hline
G=1.2 & 0.16 & 0.22 & 0.94 \\ \hline
G=1.5 & 0.06 & 0.11 & 0.93 \\ \hline
G=2 & 0.18 & 0.22 & \cellcolor[gray]{0.8} 1.09\\ \hline
\end{tabular}
\caption{Normalized variances of the 3-mode linear cluster state nullifiers, for different values of the gain.}
\label{threemodenullifier}
\end{table}

\subsection{four-mode cascaded FWM}
In the case of four-mode symmetric cascaded FWM, there are several possible graphs of cluster states. We first focus here on the linear one, whose $U_V$ matrix can also be found in \cite{Yukawa:2008iu}. Using our optimization strategy, we calculate the best possible nullifiers for different values of the gain, as shown in Table \ref{fourmodenullifier}. We see a completely different situation from the three-mode case. As the state impinging on the detectors is already an entangled state, it can be turned into a cluster state with phase controlled homodyne detection and post-processing more efficiently. In particular, we see that the values of the nullifiers follow roughly those of the squeezing values.

The same procedure can be applied to other cluster shapes, for instance we tested square and T shape clusters, which showed a very different behavior: in these cases, nullifiers values evolution is not monotonous with G values, and there is an optimal gain for each shape. Other shapes could be tested, or other types of clusters such as weighted graph \cite{Menicucci2011}. Hence, this system is readily applicable for quantum information processing. One should stress, however, that in order to exhibit cluster statistics it is necessary to precisely control the phase of the local oscillator in each homodyne detection, which can be accomplished for instance with digital locking electronics. Otherwise, it is also possible to build in the optimization routine within certain range of possible homodyne detection phase, and obtain solutions under theses constraints.

\begin{table}[h]
\centering
\begin{tabular}{|c|c|c|c|c|} \hline
FWM gain &nullifier 1 &nullifier 2 & nullifier 3 & nullifier 4\\ \hline
G=1.2 & 0.13 & 0.44 & 0.13 & 0.44 \\ \hline
G=1.5 & 0.04 & 0.25 & 0.04 & 0.25 \\ \hline
G=2 & 0.02 & 0.13 & 0.02 & 0.13\\ \hline
\end{tabular}
\caption{Normalized variances of the 4-mode linear cluster state nullifiers, for different values of the gain.}
\label{fourmodenullifier}
\end{table}

\section{summary}
In summary, we theoretically proposed to cascade two and three FWM processes to generate three-mode and four-mode cluster states respectively. The three-mode cluster state generation is sensitive to the gain values of the FWM processes. We considered the specific situation where the two FWM processes share the same gain value and found that when the gain value is below a certain value, we can construct the three-mode cluster state, but the intrinsic two mode structure of the system prevent from generating good clusters. In contrary, in the four-mode case, we found that for a wide range of gain values when the three FWM processes share the same gain value, different graphs of four-mode cluster states can be constructed. Thus, we expect that by cascading more FWM processes, multimode cluster states with different graphs can be constructed and this scheme for realizing versatile quantum networks promises potential applications in quantum information processing.

\acknowledgments
This work is supported by the European Research Council starting grant Frecquam and the French National Research Agency project Comb. Y.C. recognizes the China Scholarship Council. J. J. acknowledge the support from NSFC (Nos. 11374104 and 10974057), the SRFDP (20130076110011), the Program for Eastern Scholar at Shanghai Institutions of Higher Learning, the Program for New Century Excellent Talents in University (NCET-10-0383), the Shu Guang project (11SG26), the Shanghai Pujiang Program (09PJ1404400). X. Xu thanks the National Natural Science Foundation of China (Grant No. 11134003) and Shanghai Excellent Academic Leaders Program of China (Grant No. 12XD1402400).

\appendix*
\section{Cluster matrices}
Here are the solution for 3-mode linear cluster, with $G=1.2$:
\begin{equation}
P_{homo3-lin}=\left(
\begin{array}{ccc}
 0.52-0.86 i & 0 & 0 \\
 0 & 0.61-0.79i & 0 \\
 0& 0 & 0.93+0.36 i
\end{array}
\right)
\end{equation}

\begin{equation}
O_{post3-lin}=\left(
\begin{array}{ccc}
 0.97& -0.12 & 0.23 \\
0 & -0.88& -0.48 \\
 0.26 & 0.46 & -0.85
\end{array}
\right)
\end{equation}
The feasible cluster matrix is:
\begin{equation}
\left(
\begin{array}{ccc}
 0.21 & 0.67+0.30 i & 0.41-0.49 i \\
-0.58i & 0.30+0.49 i & -0.49+0.30 i \\
 -0.79& -0.18+0.30 i & -0.11-0.49 i
\end{array}
\right)
\end{equation}

For the 4-mode linear cluster, we find
Line shape:
The $P_{homo4-lin}$ is
\begin{equation}
\left(
\begin{array}{cccc}
 0.34-0.94 i & 0&0& 0 \\
 0 & 0.99+0.14 i & 0& 0 \\
 0& 0 & 0.19-0.98 i & 0\\
 0& 0 & 0 & 0.78-0.62 i
\end{array}
\right)
\end{equation}
The $O_{post4-lin}$ is
\begin{equation}
\left(
\begin{array}{cccc}
 0.46 & 0.15 & -0.86 & 0.17 \\
 0.20 & -0.73 & 0.11 & 0.65 \\
 0.11 & -0.65 & -0.20 & -0.73 \\
 0.86 & 0.17 & 0.46 & -0.15
\end{array}
\right)
\end{equation}
And the cluster matrix is
\begin{equation}
\left(
\begin{array}{cccc}
 -0.15-0.12 i & -0.72-0.12 i & -0.19+0.61 i & -0.16-0.04 i \\
 -0.12+0.05 i & -0.12-0.64 i & 0.61-0.09 i & -0.04+0.43 i \\
 0.20+0.60 i & 0.08-0.17 i & 0.10+0.39 i & 0.59-0.25 i \\
 0.71+0.20 i & -0.05+0.08 i & -0.22+0.10 i & -0.20+0.59 i
\end{array}
\right)
\end{equation}

\bibliographystyle{apsrev}
%\bibliography{FWM}

\begin{thebibliography}{99}
\bibitem{Braunstein:2005wr} Samuel L. Braunstein and Peter van Loock, Rev. Mod. Phys. \textbf{77,} 513 (2005).
\bibitem{Su:2007ts} Xiaolong Su, Aihong Tan, Xiaojun Jia, Jing Zhang, Changde Xie, and Kunchi Peng, Phys. Rev. Lett. \textbf{98,} 070502 (2007).
\bibitem{Yukawa:2008iu} Mitsuyoshi Yukawa, Ryuji Ukai, Peter van Loock, and Akira Furusawa, Phys. Rev. A \textbf{78} 012301 (2008).
\bibitem{Armstrong:2012tt} Seiji Armstrong, Jean-Francois Morizur, Jiri Janousek, Boris Hage, Nicolas Treps, Ping Koy Lam, and Hans-A. Bachor, Nauture Commun. \textbf{3,} 1026 (2012).
\bibitem{Pysher:2011hn} Matthew Pysher, Yoshichika Miwa, Reihaneh Shahrokhshahi, Russell Bloomer, and Olivier Pfister, Phys. Rev. Lett. \textbf{107} 030505 (2011).
\bibitem{Chen:2014jx} Moran Chen, Nicolas C. Menicucci, and Olivier Pfister, Phys. Rev. Lett. \textbf{112} 120505 (2014).
\bibitem{Yokoyama:2013jp} Shota Yokoyama, Ryuji Ukai, Seiji C. Armstrong, Chanond Sornphiphatphong, Toshiyuki Kaji, Shigenari Suzuki, Jun-ichi Yoshikawa, Hidehiro Yonezawa, Nicolas C. Menicucci and Akira Furusawa, Nature Photon. \textbf{7,} 982 (2009).
\bibitem{Roslund:2013cb} Jonathan Roslund, Renne Medeiros de Ara\'{u}ujo, Shifeng Jiang, Claude Fabre and Nicolas Treps, Nature Photon. \textbf{8,} 109 (2014).
\bibitem{Ferrini:2013cr} G Ferrini, J P Gazeau, T Coudreau, C Fabre and N Treps, New J. Phys. \textbf{15,} 093015 (2013).
\bibitem{McCormick:2007} C. F. McCormick, V. Boyer, E. Arimonda, and P. D. Lett, Opt. Lett. \textbf{32,} 178 (2007).
\bibitem{Liu:2011} Cunjin Liu, Jietai Jing, Zhifan Zhou, Raphael C. Pooser, Florian Hudelist, Lu. Zhou, and Weiping Zhang, Opt. Lett. \textbf{36,} 2979 (2011).
\bibitem{Qin:2012} Zhongzhong Qin, Jietai Jing, Jun Zhou, Cunjin Liu, Raphael C. Pooser, Zhifan Zhou, and Weiping Zhang, Opt. Lett. \textbf{37,} 3141 (2012).
\bibitem{Boyer:2008} Vincent Boyer, Alberto M. Marino, Paphael C. Pooser, Paul D. Lett, Science \textbf{321,} 544 (2008).
\bibitem{Camacho:2009} Ryan M. Camacho, Praveen K. Vudyasetu and John C. Howell, Nature Photon. \textbf{3,} 103 (2009).
\bibitem{MacRae:2012} A. MacRae, T. Brannan, R. Achal, and A. I. Lvovsky, Phys. Rev. Lett. \textbf{109,} 033601 (2012).
\bibitem{Marino:2009} A. M. Marino, R. C. Pooser, V. Boyer and P. D. Lett, Nature \textbf{457,} 859 (2009).
\bibitem{Pooser:2009} R. C. Pooser, A. M. Marino, V. Boyer, K. M. Jones, and P. D. Lett, Phys. Rev. Lett. \textbf{103,} 010501 (2009).
\bibitem{Jing:2011} Jietai Jing, Cunjin Liu, Zhifan Zhou, Z. Y. Ou and Weiping Zhang, Appl. Phys. Lett. \textbf{99,} 011110 (2011).
\bibitem{Kong:2013} Jia Kong, Jietai Jing, Hailong Wang, F. Hudelist, Cunjin Liu, and Weiping Zhang, Appl. Phys. Lett. \textbf{102} 011130 (2013).
\bibitem{Clark:2014} Jeremy B. Clark, Ryan T. Glasser, Quentin Glorieux, Ulrich Vogl, Tian Li, Kevin M. Jones and Paul D. Lett, Nature Photon. \textbf{8,} 515 (2014).
\bibitem{Qin:2014} Zhongzhong Qin, Leiming Cao, Hailong Wang, A. M. Marino, Weiping Zhang, and Jietai Jing, Phys. Rev. Lett. \textbf{113,} 023602 (2014).
\bibitem{Braunstein:2005fn} Samuel L. Braunstein, Phys. Rev. A \textbf{71,} 055801 (2005).
\bibitem{Boyd:1992} R. Boyd, \emph{Nonlinear Optics} (Academic, New York, 1992).
%\bibitem{EPR:1935} A. Einstein, B. Podolsky, and N. Rosen, Phys. Rev. \textbf{47,} 777-780 (1935).
\bibitem{vanLoock:2010kt} Ryuji Ukai, Jun-ichi Yoshikawa, Noriaki Iwata, Peter van Loock, and Akira Furusawa, Phys. Rev. A  \textbf{81,} 032315 (2010).
\bibitem{vanLoock2007} P. van Loock, Christian Weedbrook, and Mile Gu, Phys. Rev. A 76, 032321 (2007)
\bibitem{NewGiulia} G. Ferrini, J. Roslund, F. Arzani, Y. Cai, C. Fabre, and N. Treps, arXiv quant-ph, 1407, 5318. (2014).
\bibitem{Jonathan} J. Roslund, O. M. Shir, T. B\"ack, and H. Rabitz, Phys. Rev. A 80, 043415 (2009).
\bibitem{Menicucci2011} N. Menicucci, S. Flammia and P. van Loock, Phy. Rev. A \textbf{83,} 042335 (2011).

\end{thebibliography}

\end{document}